\documentclass[aps,twocolumn,amsmath,amssymb,preprintnumbers,floatfix,prb,superscriptaddress,longbibliography]{revtex4-2}

\usepackage{comment}
\usepackage[version=4]{mhchem}
\usepackage[utf8]{inputenc}
\usepackage{newtxtext}
\usepackage[upint]{newtxmath}
\usepackage{microtype}
\usepackage{textcomp}
\usepackage{dsfont}
\usepackage{eucal}
\usepackage{siunitx}
\usepackage{soul}

\usepackage{enumerate}
\usepackage{amsfonts}
\usepackage{color}
\usepackage{soul}

\usepackage{todonotes}
\presetkeys%
    {todonotes}%
    {inline}{}

\usepackage{graphicx}

\usepackage[colorlinks,allcolors=blue]{hyperref}
\usepackage[capitalize]{cleveref} 
\usepackage{cleveref}

\newcommand{\vk}{\boldsymbol{k}}

\definecolor{DarkBlue}{rgb}{0,0,0.80}
\definecolor{DarkRed}{rgb}{0.80,0,0}
\definecolor{Purple}{rgb}{0.55,0,0.55}

\begin{document}

\title{Interfacial orbital transmission, conversion, and mechanical torque in metals}
\author{Chi Sun}
\email{e0021580@u.nus.edu}
\affiliation{Aix-Marseille Université, CNRS, CINaM, Marseille, France}
\author{Dongwook Go}
\affiliation{Department of Physics, Korea University, Seoul 02841, Republic of Korea}
\author{Yuriy Mokrousov}
\affiliation{Peter Grünberg Institut (PGI-1), Forschungszentrum Jülich and JARA, 52428 Jülich, Germany}
\affiliation{Institute of Physics, Johannes Gutenberg University Mainz, 55099 Mainz, Germany}
\author{Jacob Linder}
\affiliation{Center for Quantum Spintronics, Department of Physics, Norwegian \\ University of Science and Technology, NO-7491 Trondheim, Norway}
\author{Aurélien Manchon}
\email{aurelien.manchon@univ-amu.fr}
\affiliation{Aix-Marseille Université, CNRS, CINaM, Marseille, France}

\begin{abstract}
Interfacial orbital transport remains far less understood than its bulk counterpart despite its central role in orbitronic experiments. Here, we theoretically investigate the transmission and conversion of orbital angular momentum across a metallic interface using a model Hamiltonian incorporating crystal-field effects. We show that an injected orbital dipole moment undergoes pronounced oscillations driven by the crystal field and generates characteristic quadrupole moments determined by the orbital orientation relative to the interface. Unlike spin precession, the dipole relaxes toward a finite value away from the interface. We further quantify interfacial orbital memory loss and demonstrate that orbital absorption produces a sizable mechanical torque obtained from the orbital continuity equation. 
\end{abstract}

\maketitle

\textit{Introduction. ---} Orbitronics is a recently-developed research field based on the physics of the orbital degree of freedom of electrons \cite{go2021orbitronics,jo2024spintronics,wang2025orbitronics}, which enables the design of novel low-dissipation applications including orbital memory devices \cite{yang2024orbital,ding2020harnessing,hayashi2023observation}, terahertz emitters \cite{Liu2024OrbitronicTHz,Wang2023InverseOrbital}, and logic circuits \cite{wang2025orbitronics}. Non-equilibrium orbital moments can manifest in metals in the absence of spin-orbit coupling, and are therefore present in light, cheap, and abundant metals, largely free from scarce materials such as heavy metals and rare-earth metals \cite{salemi}. As a result, the development of orbitronics plays an essential role in achieving sustainable and resilient advanced microelectronics. 

Orbital transport in metals has attracted significant attention lately \cite{go2021orbitronics}, because some preliminary results have shown that orbital transport presents several key advantages compared to spin transport, e.g., much larger orbital Hall effect governed by the crystal field rather than spin-orbit coupling in a wide range of transition metals \cite{jo2018gigantic,kontani2009giant,go2018intrinsic}, potentially longer orbital diffusion length of the orbital Hall materials than
the spin diffusion length of the spin Hall materials \cite{Ti_nature,hayashi2023observation,PhysRevApplied.15.L031001,hotspots}, larger orbital Edelstein effect compared with spin Edelstein effect in diverse noncentrosymmetric systems \cite{Salemi2019OrbitallyDominated,PhysRevResearch.3.013275}, and so on. Whereas experimental evidence of orbital transport continues to accumulate, the propagation of the orbital moment across metallic interfaces remains poorly understood. Indeed, most theoretical models so far focus on the orbital transport properties in the bulk of metals \cite{jo2018gigantic,go2018intrinsic,hotspots,Han,pezo,orbital_Kerr,Han2023OrbitalTextures}, disregarding the interfacial effects, such as transmission and memory loss. Due to the complexity of orbital hybridization across an interface and the high sensitivity of orbital moments to the chemical environment, such interfacial effects are crucial to develop phenomenological models of orbital transport and interpret experiments. This will also open possibilities to broadly tune the orbital propagation by interface engineering. 

To clarify the interfacial effects, we consider a minimal model of a bilayer composed of two semi-infinite free-electron layers, one of them accommodating a spherical crystal field describing essential features of the orbital textures. This model can be readily extended to explore more complex spin and orbital features. We theoretically investigate the response of a metal with crystal-field-induced orbital texture to the injection of orbital dipole moments from an adjacent metal. The transmission of dipole moments is accompanied by the generation of specific quadrupole moments, determined by the orientation of the injected orbital moment and by the crystal field. While coupled oscillations between dipole and quadrupole moments are a generic feature of bulk orbital dynamics, our model clarifies which quadrupole components are permitted across a metallic interface. Interfacial effects—including orbital memory loss, dipole-to-quadrupole conversion efficiency, and the orbital Rashba effect—are explicitly addressed. Finally, our results indicate that the mechanical torque that results from angular momentum transfer to the lattice can be remarkably large, highlighting the potential of orbital transport phenomena for torque-based applications.

\begin{figure}
\includegraphics[width=\columnwidth]{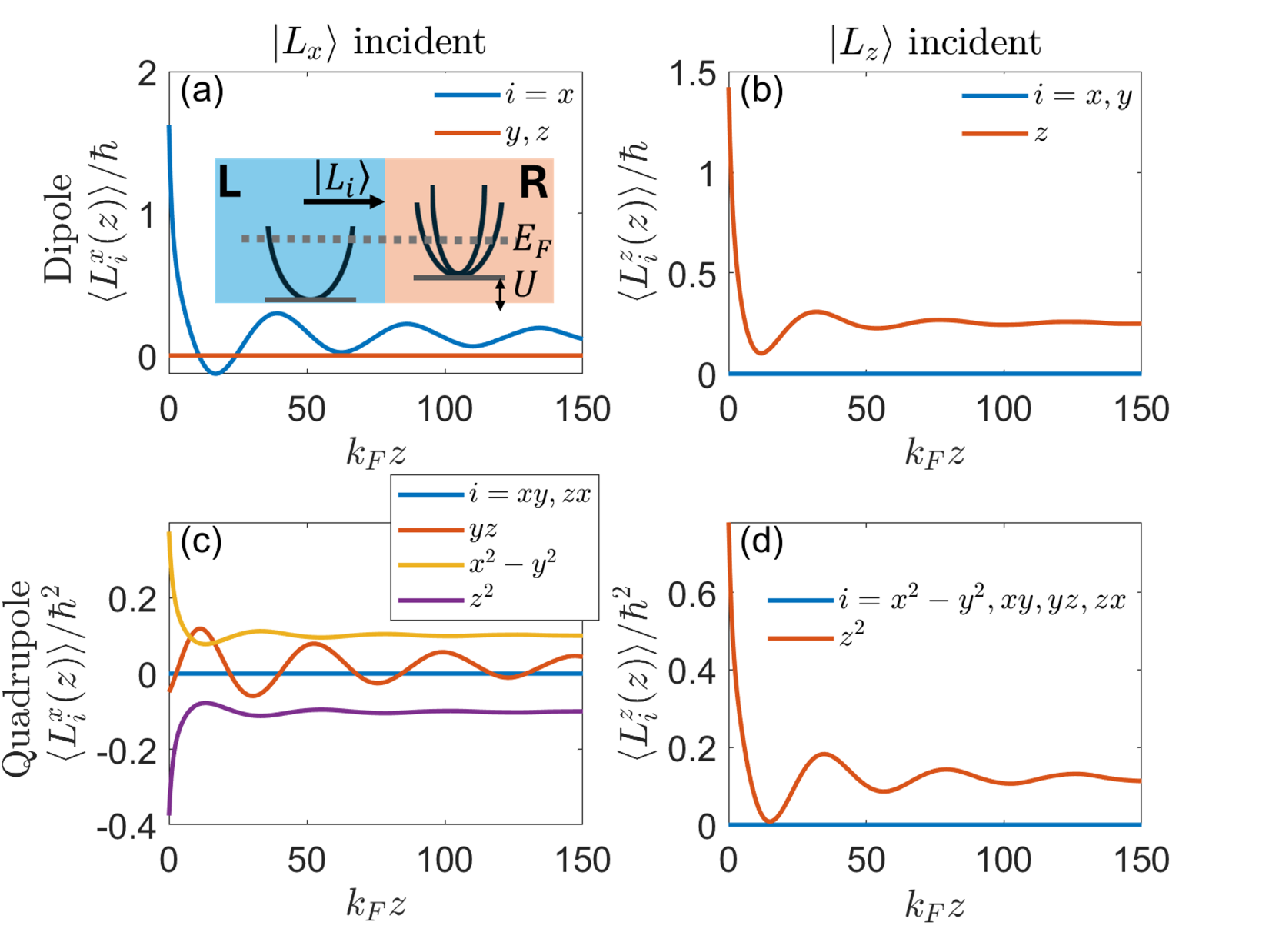}
	\caption{(a-b) Orbital dipole and (c-d) quadrupole moments as a function of $z$ in the right layer induced by incident $|L_x\rangle$ and $|L_z\rangle$ dipole moments from the left layer. The short-handed notation $\langle L_i^j\rangle$ represents $\langle L_i\rangle=\langle \Psi_\text{R}|\boldsymbol{L}_i|\Psi_\text{R}\rangle$ induced by incident $|L_j\rangle$. The parameters used are: $U/E_F =0.5$ and $r/(1/2m)=0.5$, which gives $k_t/k_F=\sqrt{1/3}\approx0.577$ and $k_r/k_F=\sqrt{1/2}\approx0.707$. Here $\alpha_R^I=0$. The inset in (a) represents a schematic diagram of the bilayer structure with energy bands.}
    \label{fig:L_all}
\end{figure}

\textit{Theory. ---} To study the orbital transmission and conversion across a metallic interface, we consider a bilayer structure [see inset of Fig. \ref{fig:L_all}(a)] in which the orbital moment is created and injected from the left layer into the right layer with intrinsic orbital textures. A free electron model expressed by $H_\text{L}=\hbar^2\vk^2/2m$ is employed for the left layer. The right layer accepting the orbital injection is described by \cite{Bernevig2005Aug}
\begin{align}
H_\text{R}&=\frac{\hbar^2\boldsymbol{k}^2}{2m}+U+r(\boldsymbol{L}\cdot\boldsymbol{k})^2,\label{eq:H_R}
\end{align}
where $U$ is the local chemical potential difference (compared with the left layer). The essential feature of the Hamiltonian
is the coupling between the local orbital moment $\boldsymbol{L}$ and the crystal momentum $\boldsymbol{k}$, where $r$ is the coupling strength that gives the intrinsic orbital texture. This term corresponds to a centrosymmetric crystal field, which can be applied to various materials, including $p$-doped silicon and graphane \cite{Bernevig2005Aug,p_dope_graphene}, $sp$ metals \cite{go2018intrinsic}, and $d$ transition metals \cite{jo2018gigantic}. Here we consider a $p$-orbital system with the orbital angular momentum operator matrices $\boldsymbol{L}=(\boldsymbol{L}_x,\boldsymbol{L}_y,\boldsymbol{L}_z)$ given by
    \begin{equation}
    \boldsymbol{L}_x=\frac{\hbar}{\sqrt{2}}\begin{pmatrix}
0 & 1 & 0\\
1 & 0 & 1\\
0 & 1 & 0
\end{pmatrix},  \boldsymbol{L}_y=\frac{\hbar}{\sqrt{2}i}\begin{pmatrix}
0 & 1 & 0\\
-1 & 0 & 1\\
0 & -1 & 0
\end{pmatrix}, \boldsymbol{L}_z=\hbar\begin{pmatrix}
1 & 0 & 0\\
0 & 0 & 0\\
0 & 0 & -1
\end{pmatrix}
\label{eq:L_basis}
\end{equation}
in the angular momentum basis $|m_l=+1\rangle=(1,0,0)^T$, $|m_l=0\rangle=(0,1,0)^T$ and $|m_l=-1\rangle=(0,0,1)^T$. Since the orbital quantization axis is along $\hat{z}$, it is convenient to choose the interface normal to $\hat{z}$ without loss of generality. We then focus on the transport along $\hat{z}$ by utilizing a cylindrical system of coordinates  with the propagation wave vector $k_z$ and the conserved transverse components $k_x=\kappa\cos\phi$ and $k_y=\kappa\sin\phi$, thus satisfying $k^2=\kappa^2+k_z^2$ with $k=|\vk|$. Note that the centrosymmetric or spherical system reduces to cylindrical due to the interface, which breaks symmetry.

Next we build the wavefunctions. In the left layer, the orbital dipole moments being injected into the right layer are given by $|L_x\rangle = \frac{1}{2}|m_l=+1\rangle + \frac{1}{\sqrt{2}}|m_l=0\rangle + \frac{1}{2}|m_l=-1\rangle$, $|L_y\rangle = \frac{1}{2}|m_l=+1\rangle + \frac{i}{\sqrt{2}}|m_l=0\rangle - \frac{1}{2}|m_l=-1\rangle$ and $|L_z\rangle= |m_l=+1\rangle$, thus satisfying $\langle L_i|\boldsymbol{L}_j|L_i\rangle=\hbar\delta_{ij}$. Given the injections, the wavefunction of the left layer is written as
\begin{align}
    \Psi_\text{L}=|L_i\rangle e^{ik_z^N z}+e^{-ik_z^N z}\big(&r_{-1}|m_l=-1\rangle +r_{0}|m_l=0\rangle\notag\\&+r_{+1}|m_l=+1\rangle\big),
\end{align}
in which $i=x,y,z$ describes different orbital dipole injections. The $z$-component of the wave vector $k_z^N=\sqrt{k_F^2-\kappa^2}$ with $k_F=\sqrt{2mE_F}/\hbar$ being the Fermi wave vector. $r_{-1,0,+1}$ represents the corresponding reflection coefficients. 

To get the wavefunction of the right layer, we note that the helicity $\lambda=\boldsymbol{k}\cdot\boldsymbol{L}/{k}$ is a good quantum number of $H_\text{R}$. As a result, $H_\text{R}$ shares common eigenstates of the orbital helicity operator $\lambda$ given by
\begin{align}
|\lambda=+\hbar\rangle&=\frac{((k_z+k)^2,\sqrt{2}(k_z+k)\kappa e^{i\phi},\kappa^2 e^{2i\phi})^T}{(k_z+k)^2+\kappa^2},\notag\\ |\lambda=-\hbar\rangle&=\frac{((k_z-k)^2,\sqrt{2}(k_z-k)\kappa e^{i\phi},\kappa^2 e^{2i\phi})^T}{(k_z-k)^2+\kappa^2},\\|\lambda=0\rangle&=\frac{(-\kappa e^{-i\phi},\sqrt{2}k_z,\kappa e^{i\phi})^T}{\sqrt{2}k},\notag 
\end{align}
in which $|\lambda=\pm\hbar\rangle$ has the same eigen-energy $\epsilon_k^{t}=(1/2m+r)\hbar^2k^2$ and $|\lambda=0\rangle$ has eigen-energy $\epsilon_k^{r}=\hbar^2k^2/2m$. In other words, the energy bands contain two degenerate bands of helicity
$\lambda = \pm\hbar$ as well as a third band of helicity $\lambda = 0$. Based on the eigenstates and energy bands, the wave function of the right layer is given by
\begin{align}
    \Psi_\text{R}=&e^{ik_z^t z}t_{+1}|\lambda=+\hbar\rangle_{k_z=k_z^t}+e^{ik_z^t z}t_{-1}|\lambda=-\hbar\rangle_{k_z=k_z^t}\notag\\&+e^{ik_z^r z}t_0|\lambda=0\rangle_{k_z=k_z^r},
\end{align}
where $t_{-1,0,+1}$ denotes the corresponding transmission coefficients. Note that different energy bands give different $k_z$ as $k_z^t=\sqrt{k_t^2-\kappa^2}$ with $k_t=\sqrt{2m(E_F-U)/(1+2mr)}/\hbar$, and $k_z^r=\sqrt{k_r^2-\kappa^2}$ with $k_r=\sqrt{2m(E_F-U)}/\hbar$, which are obtained from solving $\epsilon_k^{t}+U=\epsilon_k^{r}+U=E_F$ since we are interested in conduction electrons. The difference between $k_z^t$ and $k_z^r$ gives rise to oscillation of the orbital transport, as will be discussed later. The wavefunctions above are used to describe the transport of the {\em atom-centered} orbital angular momentum. The orbital angular momentum arising from the self-rotation of an electron about its center of mass —which is particularly enhanced in non-centrosymmetric systems— is not included in our centrosymmetric model.

To determine the reflection and transmission coefficients in $\Psi_\text{L}$ and $\Psi_\text{R}$, we need to apply the boundary conditions at the interface. The interfacial orbital Rashba effect can also be included by using the interface Hamiltonian $H_\text{I}=\alpha_R^I\boldsymbol{L}\cdot(\boldsymbol{k}\times\hat{z})\delta(z)$ with $\alpha_R^I$ being the orbital Rashba strength. 
By combining $H_\text{L}$, $H_\text{R}$, and $H_\text{I}$, the boundary conditions are derived as the continuity of the wavefunction and the discontinuity of its first derivative introduced by the crystal field strength $r$ in the right layer and the interfacial Rashba $\alpha_R^I$ strength (see Supplemental Material for details \cite{SuppMat}). As discussed below, it is found that the discontinuities introduced by $r$ and $\alpha_R^I$ play essential but different roles for the interfacial transport.

\textit{Orbital dipole and quadrupole moment injection.--} At equilibrium, there is intrinsic orbital texture but no net orbital polarization in the right layer. The orbital injection drives the system out-of-equilibrium. Here, we investigate the injection of orbital dipole moments from the left layer into the right one. To do so, we define the observable $\langle L_i^j\rangle$ as the $i$-th orbital moment component induced by the injection of the $j$-th orbital moment from left to right, i.e., $\langle L_i^j\rangle=\langle \Psi_\text{R}|\boldsymbol{L}_i|\Psi_\text{R}\rangle$ when $|L_j\rangle$ is incident. The averaged orbital density is evaluated as $\langle L_i^j(z)\rangle=\frac{1}{\pi k_F^2}\int_0^{2\pi}d\phi \int_0^{k_F}\kappa d\kappa\langle L_i^j (\kappa,z,\phi)\rangle$, where different (transverse) incident angles (determined by $\kappa$ and $\phi$) are integrated and then averaged over the Fermi surface.

We first consider the orbital dipole operators $\boldsymbol{L}_{x,y,z}$ given in Eq. (\ref{eq:L_basis}). As shown in Fig. \ref{fig:L_all}(a) with a finite crystal field [$r/(1/2m)=0.5$], only $\langle L_x^x\rangle$ survives with a damped oscillatory behavior while $\langle L_y^x\rangle$ and $\langle L_z^x\rangle$ remain zero. This is different from spin precession, where the oscillation of the $x$-component generates nonzero $y$ and $z$-components \cite{anatomy_STT}. The oscillation originates from the crystal field $r$, which gives different propagation wavevectors $k^t_z$ and $k_z^r$. The averaged damped behavior can be explained as follows \cite{SuppMat}. For small $\kappa$, stationary oscillation with constant amplitude of $\langle L_i^x(\kappa,z,\phi)\rangle$ with $i=x,y,z$ develops. As $\kappa$ increases and exceeds $k_t$ or $k_r$, $k_z$ becomes imaginary and gives decaying behavior. Similar behaviors are obtained for incident $|L_y\rangle$ and $|L_z\rangle$, where only the orbital dipole moments with the same orientation as the incident one can be transmitted, while the transverse ones remain zero, indicating no conversion between different dipole moments. Moreover, $\langle L_y^y\rangle$ (not shown here) is exactly the same as $\langle L_x^x\rangle$, due to cylindrical symmetry.

Interestingly, we find that the oscillations of the orbital dipole moment are accompanied by the generation of specific quadrupole moments. Here we use the short-hand notations for the torsional quadrupole operators $\boldsymbol{L}_{xy}=\frac{1}{2} \{\boldsymbol{L}_x,\boldsymbol{L}_y \}$, $\boldsymbol{L}_{yz}=\frac{1}{2} \{\boldsymbol{L}_y,\boldsymbol{L}_z \}$ and $\boldsymbol{L}_{zx}=\frac{1}{2} \{\boldsymbol{L}_z,\boldsymbol{L}_x \}$, which are related to the orbital torsion \cite{Han} and satisfy $\langle L_i|\boldsymbol{L}_j|L_i\rangle=0$ with $j=xy,yz,zx$ for incident dipole moments $i=x,y,z$. In Fig. \ref{fig:L_all}(c), it is seen that the oscillating torsional quadrupole $\langle L_{yz}^x
\rangle$ (red curve) is generated in response to the $|L_x\rangle$ injection, implying the conversion between dipole and torsional quadrupole moments. The coupled oscillation between orbital dipole and quadrupole moments is a key feature of orbital dynamics \cite{Han}. However, as shown in Fig. \ref{fig:L_all}(d) for incident $|L_z\rangle$ (along the orbital quantization direction), none of the torsional quadrupoles can be generated. The different behaviors induced by incident $|L_x\rangle$ and $|L_z\rangle$ can be understood by the corresponding commutation relations describing the interaction between the (injection) orbital operator and the crystal field [see Eq. \eqref{eq:cont} below]:
\begin{eqnarray}
\frac{d\langle \boldsymbol{L}_x\rangle}{dt}&\sim&\frac{1}{i\hbar}\langle[\boldsymbol{L}_x,(k_x\boldsymbol{L}_x+k_y\boldsymbol{L}_y)^2]\rangle=2\langle k_y^2\boldsymbol{L}_{yz}\rangle,\\
\frac{d\langle \boldsymbol{L}_z\rangle}{dt}&\sim&\frac{1}{i\hbar}\langle[\boldsymbol{L}_z,(k_x\boldsymbol{L}_x+k_y\boldsymbol{L}_y)^2]\rangle=0,
\end{eqnarray}
where $\langle...\rangle$ stands for averaging over the Fermi surface. 

Besides the torsional quadrupoles, the other two quadrupole components are given by $\boldsymbol{L}_{x^2-y^2}=\frac{1}{2}(\boldsymbol{L}_x^2-\boldsymbol{L}_y^2)$ and  $\boldsymbol{L}_{z^2}=\frac{1}{2}(3\boldsymbol{L}_z^2-\boldsymbol{L}^2)$, and measure the orbital polarization \cite{Han}. Notably, the polarization quadrupoles give  $\langle L_i|\boldsymbol{L}_j|L_i\rangle\neq0$ with $j=x^2-y^2,z^2$ for incident dipole moments $i=x,y,z$. This implies that dipole injection is intrinsically associated with the polarization quadrupole and cannot be generated alone, in contrast to the spin angular momentum. As a result, nonzero polarization quadrupoles appear in Fig. \ref{fig:L_all} (c) and (d) for different injected dipole orientations. The injection of $|L_x\rangle$ ($|L_y\rangle$) gives antisymmetric outputs for $\langle L_{x^2-y^2}^{x(y)}\rangle$ and $\langle L_{z^2}^{x(y)}\rangle $, while the injection of $|L_z\rangle$ is only accompanied by the polarization quadrupole $\langle L_{z^2}^z\rangle$.

\begin{figure}
\includegraphics[width=\columnwidth]{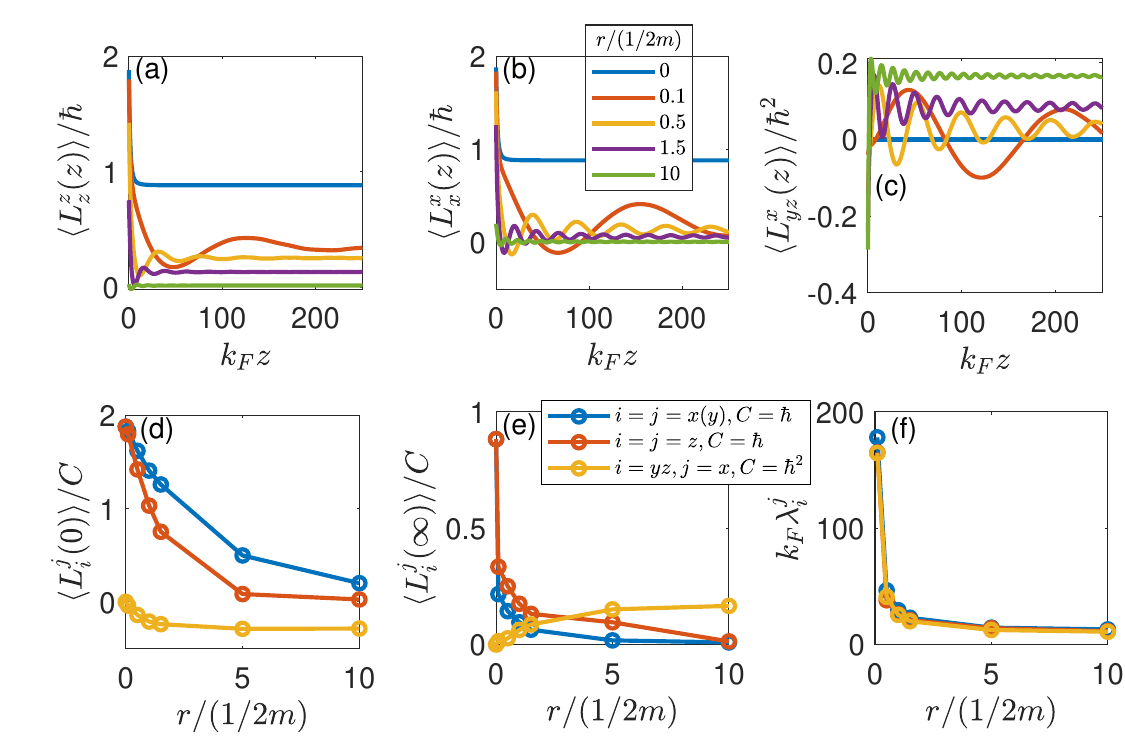}
	\caption{First row: (a-b) Orbital dipole moments and (c) quadrupole moments as a function of $z$ for different $r$ values. Second row: The crystal field $r$-dependence of the orbital dipole and quadrupole moments at (d) $z=0$, (e) $z=\infty$, and (f) their oscillation wavelengths. The other parameters are the same as used in Fig. \ref{fig:L_all}.}
    \label{fig:r_dep_all}
\end{figure}

We now investigate the crystal field $r$ dependence of the resulting dipole and quadrupole moments in Fig. \ref{fig:r_dep_all}. In the limit $r=0$, the crystal field in the right layer vanishes and the orbital dipole moments (blue curves) in Figs. \ref{fig:r_dep_all}(a,b) decay fast to a finite constant value right after leaving the interface at $z=0$ without oscillations. This decay to a finite value originates from the nonzero $U=0.5E_F$, which gives an imaginary propagation wavevector $k_t^z=k_r^z$ for large $\kappa$ since $k_r=k_t<k_F$. As $r$ increases, oscillations with higher frequencies appear and the orbital moments finally approach zero for large $z$ when $r$ is large enough, corresponding to increasing orbital angular momentum transferred into the crystal as well as a shorter orbital relaxation length. On the other hand, the torsional quadrupole moment, e.g., $\langle L_{yz}^x\rangle$ in Fig. \ref{fig:r_dep_all}(c), remains zero in the absence of $r$. Conversely, it develops with oscillations as $r$ increases and reaches a finite long-range value for large $z$. As shown in the second row of Fig. \ref{fig:r_dep_all}, the interfacial transmission of the dipoles, their long-range value, and their oscillation wavelength all vanish upon increasing $r$. In contrast, the long-range value of the torsional quadrupole increases with $r$ since this type of quadrupole is generated from the interaction with the crystal field $r$. As for the polarization quadrupole moments $\boldsymbol{L}_{x^2-y^2}$ and $\boldsymbol{L}_{z^2}$, they share similar trends as the dipole moments with respect to $r$ \cite{SuppMat}, which corresponds to their intrinsic coupling with the injected dipoles. Finally, it is found that increasing the interfacial orbital Rashba coupling $\alpha_R^I$ slightly decreases the amplitude of both dipole and quadrupole moments injected into the right layer \cite{SuppMat}.

\textit{Interfacial orbital memory loss and conversion efficiency.---}
In addition to the oscillatory decay of the orbital moment and the generation of quadrupole moments due to the crystal field in the right layer, it is important to stress out that, at the very interface, the orbital moment experiences memory loss and conversion. This memory loss emerges due to the boundary conditions derived in Supplemental Material \cite{SuppMat}. Since the crystal field couples $\boldsymbol{k}$ and $\boldsymbol{L}$, the discontinuity of the crystal field at the interface creates a discontinuity in the wave function's first derivative. Meanwhile, the interfacial orbital Rashba effect also creates a discontinuity. As for the spin counterpart, it is known that spin memory loss only occurs in the presence of spin-orbit coupling or magnetic impurities. 

Similar to the interfacial spin memory loss \cite{SML,liu2022calculating}, we derive the orbital (dipole) memory loss parameter $\delta$ by analytically solving the drift-diffusion equations for the orbital accumulations (or moments) $\mu_{\text{L,R,I}}$ and orbital dipole currents $J_{\text{L,R,I}}$ in the left (L), right (R) layers and interface (I) \cite{SuppMat}. 
Here, the L/R interface is initially considered as a bulk-like material
with orbital resistivity $\rho_\text{I}$, orbital diffusion length $l_\text{I}$, and thickness $t_\text{I}$. The continuity of orbital accumulation and current is then applied as boundary conditions at the L/I and R/I interfaces. By taking the limit $t_\text{I}\rightarrow0$, the areal interface resistance $AR_\text{I}$ and the orbital memory loss
parameter $\delta$ are defined as \cite{SML,liu2022calculating} 
\begin{equation}
   AR_\text{I}\equiv\lim_{t_\text{I}\rightarrow0}\rho_{\text{I}},\quad \delta\equiv\lim_{t_\text{I}\rightarrow0}t_\text{I}/l_\text{I},
\end{equation}
which satisfies
\begin{equation}
    \frac{J_\text{L}(z=0)}{J_\text{R}(z=0)}=\cosh{\delta}+\frac{\rho_\text{R}l_\text{R}}{AR_\text{I}}\delta\sinh{\delta}.\label{eq:JLR}
\end{equation}
$\rho_\text{R}$ and $l_\text{R}$ represent the orbital resistivity and diffusion length of the right layer, respectively. By inserting the interfacial current values into Eq. (\ref{eq:JLR}), $\delta$ can be solved numerically. 

\begin{figure}
\includegraphics[width=0.85\columnwidth]{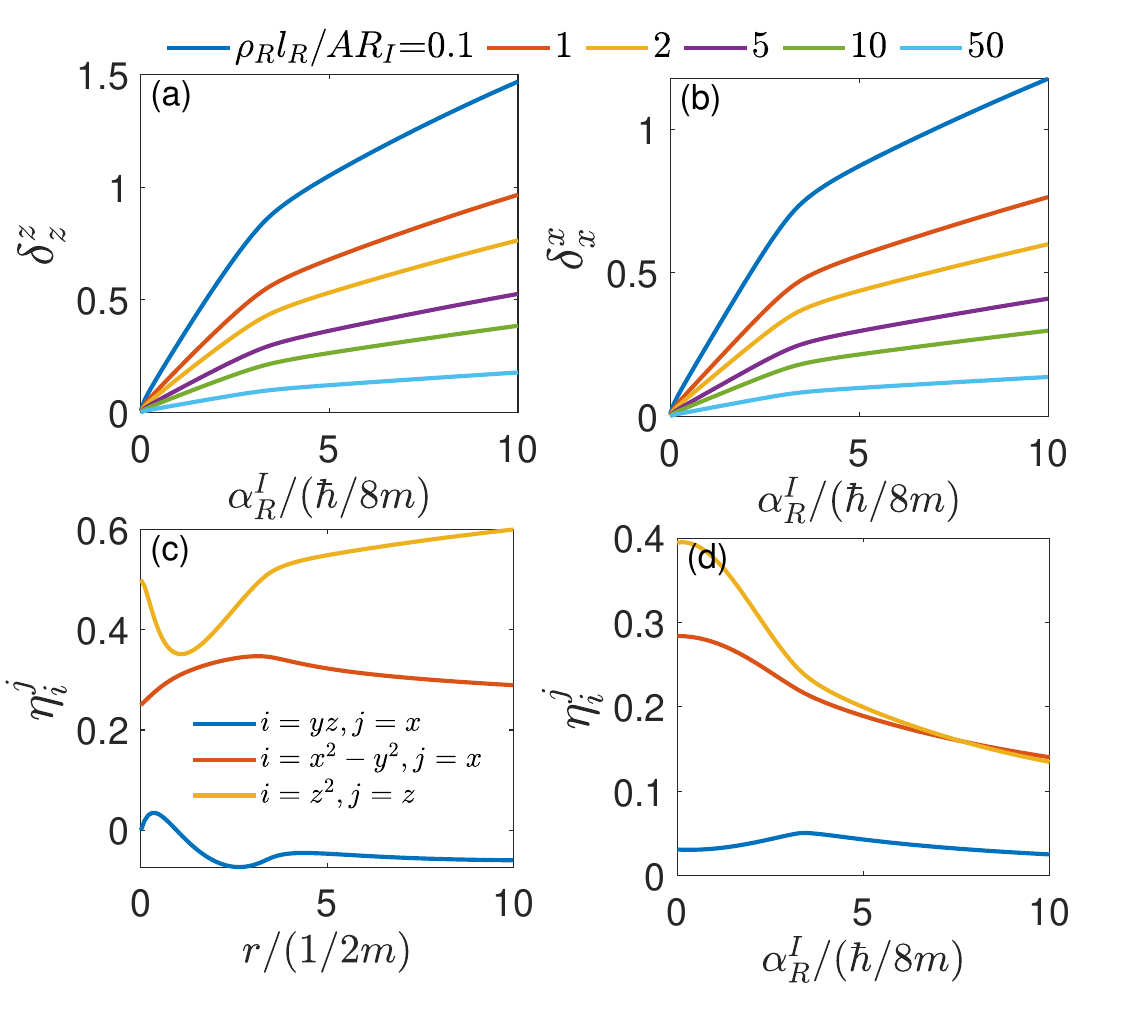}
	\caption{
    (a-b) Interfacial orbital dipole memory loss $\delta$ as a function of the interfacial Rashba strength $\alpha_R^I$. Dipole to quadrupole interfacial conversion efficiency $\eta$ as a function of the crystal field strength $r$ and $\alpha_R^I$ in (c) and (d), respectively, in which $\alpha_R^I=0$ is used in (c) while $r/(1/2m)=0.5$ is considered in (d). The other parameters are the same as used in Fig. \ref{fig:L_all}.}
    \label{fig:SML}
\end{figure}

Let us consider an incident $|L_z\rangle$ state as an example. In the left (right) layer, the orbital dipole current can be calculated as $J_{\text{L}(\text{R})}=\langle\Psi_{\text{L}(\text{R})}|\boldsymbol{J}_{L_z}|\Psi_{\text{L}(\text{R})}\rangle$ with the orbital current operator $ \boldsymbol{J}_{L_z}=\frac{1}{2}\{\boldsymbol{L}_z,\boldsymbol{v}\}$ in terms of the velocity operator $\boldsymbol{v}_{\text{L(R)}}=\frac{1}{\hbar}\partial H_{\text{L(R)}}/\partial k_z$. Explicitly, we have $\boldsymbol{v}_{\text{L}}=\frac{\hbar k_z}{m}$ and $\boldsymbol{v}_{\text{R}}=\boldsymbol{v}_{\text{L}}+r\{\vk\cdot\boldsymbol{L},\boldsymbol{L}_z\}/\hbar$ with $k_z=-i\nabla_z$. In the absence of $\alpha_R^I$, it is found that the interfacial discontinuities in the first derivative of wavefunctions due to $r$ do not result in discontinuities of the orbital dipole moment as well as current, giving $J_\text{L}(z=0)=J_\text{R}(z=0)$, and leading to zero orbital memory loss $\delta=0$. Conversely, discontinuities introduced by the interfacial orbital Rashba term, $\alpha_R^I$, produce orbital memory loss. In Fig. \ref{fig:SML}, the orbital memory loss $\delta$ is plotted as a function of $\alpha_R^I$ for incident $|L_z\rangle$ and $|L_x\rangle$, for different $\rho_\text{R}l_\text{R}/AR_\text{I}$ values in (a) and (b), respectively. Typical values $\rho_\text{R}=10$ $(\hbar/e)\mu\Omega\cdot\text{cm}$, $l_\text{R}=10$ nm and $AR_\text{I}=0.5$ m$(\hbar/e)\Omega\cdot\mu\text{m}^2$ give $\rho_\text{R}l_\text{R}/AR_\text{I}=2$, which characterizes the transparency. To enhance the orbital dipole transmission with less interfacial orbital memory loss $\delta$, it is found that smaller $\alpha_R^I$ and larger $\rho_\text{R}l_\text{R}/AR_\text{I}$ are preferred. This is unlike the spin counterpart, in which a tunnel barrier with large $AR_I$ is used to maximize the spin injection with less $\delta$. This could be attributed to different origins of the interfacial discontinuities.

On the other hand, we investigate the interfacial conversion efficiency $\eta= Q_\text{R}(z=0)/\hbar J_\text{L}(z=0)$ between different dipole and quadrupole moments. $Q$ denotes the orbital quadrupole current, and the additional $\hbar$ is included to cover the unit difference between dipole and quadrupole. Take incident $|L_x\rangle$ as an example, which is accompanied with the orbital dipole current $J_\text{L}=\langle\Psi_\text{L}|\boldsymbol{J}_{L_x}|\Psi_\text{L}\rangle$ with $ \boldsymbol{J}_{L_x}= \frac{1}{2}\{\boldsymbol{L}_x,\boldsymbol{v}_\text{L}\}$. To get its conversion to the torsional quadrupole measured by $\boldsymbol{L}_{yz}$, the quadrupole current is computed as $Q_\text{R}=\langle\Psi_\text{R}|\boldsymbol{Q}_{yz}|\Psi_\text{R}\rangle$ with $\boldsymbol{Q}_{L_{yz}}=\frac{1}{2}\{\boldsymbol{L}_{yz},\boldsymbol{v}_\text{R}\}$.  As shown by the blue curve in Fig. \ref{fig:SML}(c), the interfacial conversion efficiency $\eta_{yz}^x$ increases from zero as the crystal field $r$ and reaches its maximum, after that it decreases to zero and changes sign with $r$. 
The non-monotonic trend is also different from the monotonic behavior of the $\langle L_{yz}^x(z=0)\rangle$ as a function of $r$ [yellow curve in Fig. \ref{fig:r_dep_all}(d)]. On the other hand, the interfacial conversion efficiency to the polarization quadrupole [yellow and red curves in Fig. \ref{fig:SML}(c)] starts from a finite value rather than zero in the absence of $r$, which can be explained by the intrinsic association of the polarization quadrupole with the dipole injection, as discussed before. As $r$ increases, non-monotonic behaviors are also shown, indicating additional conversion between dipole and polarization quadrupole due to $r$, besides the intrinsic polarization quadrupole injection. In Fig. \ref{fig:SML}(d) with $r/(1/2m)=0.5$, the dipole to torsional quadrupole conversion efficiency $\eta_{yz}^x$ maximum can be achieved by modulating $\alpha_R^I$. Conversely, the polarization quadrupole counterparts $\eta_{x^2-y^2}^x$ and $\eta_{z^2}^z$ decrease as $\alpha_R^I$, indicating the quenching of the orbital polarization at the interface by the orbital Rashba $\alpha_R^I$.

\begin{figure}
\includegraphics[width=\columnwidth]{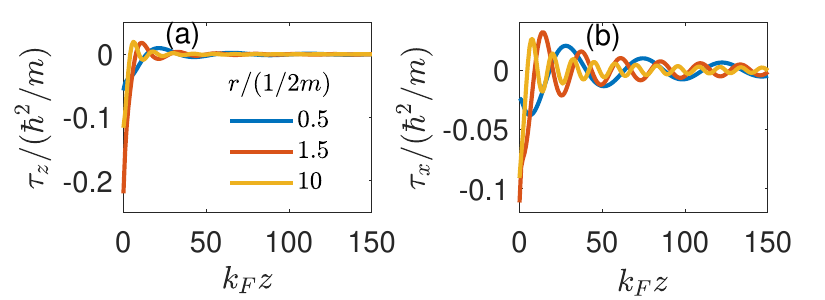}
	\caption{The orbital torque induced by (a) $|L_z\rangle$ and (b) $|L_x\rangle$ injections as a function of $z$ for different crystal field $r$, respectively. The other parameters are the same as used in Fig. \ref{fig:L_all}.}
    \label{fig:torque}
\end{figure}

\textit{Continuity equation and mechanical torque.--} The absorption of the orbital moment by the lattice, mediated by the crystal field, induces a mechanical torque. Here, we evaluate the orbital torque generated in the right layer induced by the orbital dipole injections. Using the Ehrenfest Theorem for the orbital dipole operator $\boldsymbol{L}_z$ (with incident $|L_z\rangle$), we have 
\begin{equation}\label{eq:cont}
\frac{d\langle \boldsymbol{L}_z \rangle}{dt}=\frac{1}{i\hbar}\langle [\boldsymbol{L}_z,\frac{\hbar^2\vk^2}{2m}]\rangle+\frac{1}{i\hbar}\langle [\boldsymbol{L}_z,r(\vk\cdot\boldsymbol{L})^2]\rangle,
\end{equation}
in which $\langle \partial \boldsymbol{L}_z/\partial t\rangle=0$ is applied since the operator $\boldsymbol{L}_z$ is time-independent. Inserting the orbital dipole current as $ \boldsymbol{J}_{L_z}=\langle \{\boldsymbol{L}_z,\boldsymbol{v}_\text{R}\}\rangle/2$, we arrive at the continuity equation $d\langle \boldsymbol{L}_z \rangle/{dt}=-\nabla_z \boldsymbol{J}_{L_z}+\tau_z.$
Numerically, it is found $d\langle \boldsymbol{L}_z \rangle/dt=0$, which corresponds to the steady state. This gives the orbital torque (crystal-field or lattice torque) $\tau_z$ satisfying 
\begin{equation}
    \tau_z= \langle [\boldsymbol{L}_z,r(k_x\boldsymbol{L}_x+k_y\boldsymbol{L}_y)^2]\rangle/i\hbar=\nabla_z \boldsymbol{J}_{L_z},
\end{equation} 
where the $k_z$-dependent components in $\langle[\boldsymbol{L}_z,r(\vk\cdot\boldsymbol{L})^2]\rangle/i\hbar$ are absorbed into the current divergence with $k_z=-i\nabla_z$. $\tau_z=\nabla_z\boldsymbol{J}_z$ is in a fashion reminiscent of the spin transfer torque which is determined by the divergence of the spin current \cite{SLONCZEWSKI1996L1,anatomy_STT,Chi_field_free}. As the orbital current is conserved in the normal metal, but changes as it enters the region with the crystal field $r$, there is a transfer of angular momentum to the lattice. Here, we estimate the mechanical torque exerted on the lattice by using the torque term $\tau_z$ in the continuity equation derived above.
As shown in Fig. \ref{fig:torque}(a), the orbital torque $\tau_z$ oscillates and finally damps to zero. 

Based on the blue curve in Fig. \ref{fig:torque}(a) with moderate crystal field $r/(1/2m)=0.5$, we numerically examine the "averaged" $z$-independent torque as $\bar{\tau}_z=\frac{1}{k_FD}\int_0^{k_F z=+\infty}\tau_z d(k_F z)\approx-\frac{{0.3}\hbar^2}{k_FDm}$ with the unit of Joule. Here $D$ denotes the thickness of the right layer. Consider a cylinder made of aluminum with radius $R=1$ nm and height 10 nm, its mass can be obtained as $M=8.48\times10^{-23}$ kg. Consequently, its moment of inertia can be calculated as $I=\frac{1}{2}MR^2$. By using the orbital torque $\bar{\tau}_z$ to rotate the cylinder as a mechanical torque, the angular acceleration is calculated as $\dot{\omega}_z=\frac{(\bar{\tau_z}\pi k_F^2) \pi R^2}{I}=\frac{-{0.3}(\hbar^2/m)\pi^2k_F^2 R^2}{\frac{1}{2}k_FDMR^2}\approx-{870}$ $\text{rad}/\text{s}^2$ for $D=10$ nm, in which $\bar{\tau}\pi k_F^2$ with $k_F=10^{10}$ m$^{-1}$ gives the torque density per area. For a larger $r/(1/2m)=1.5$, the angular acceleration increases to $\dot{\omega}_z\approx-{1770}$ $\text{rad}/\text{s}^2$. However, it decreases to $-{720}$ $\text{rad}/\text{s}^2$ when $r/(1/2m)=10$, {indicating a non-monotonic behavior regarding $r$.} As for the $|L_x\rangle$-induced torques shown in Fig. \ref{fig:torque}(b), we obtain the angular accelerations $-{900}$ $\text{rad}/\text{s}^2$, $-{1140}$ $\text{rad}/\text{s}^2$ and $-{580}$ $\text{rad}/\text{s}^2$ for $r/(1/2m)=0.5$, 1.5 and 10, respectively. Our simplified calculation indicates that a huge mechanical torque is achievable by injecting orbital dipole moments into a metal with a moderate crystal field $r$ from an adjacent metal. Further \textit{ab initio} calculations are desired to quantitatively estimate the torque in realistic materials. Besides the mechanical rotation, the absorbed orbital moment can also be dissipated in other ways, \textit{e.g.}, exciting chiral phonons. 

\textit{Concluding remarks---} 
Before concluding, we briefly comment on possible experimental routes to detect the orbital dipole and quadrupole moments. One approach relies on electrical detection via their conversion into transverse charge currents through the inverse orbital Hall effect for dipole moments \cite{Wang2023InverseOrbital,Xu2024Orbitronics} and the inverse orbital-torsion Hall effect for quadrupole moments \cite{Han}. Alternatively, since orbital quadrupole moments induce anisotropies in orbital occupation and orbital mixing, they may be probed through spatially resolved or anisotropy-sensitive techniques, such as resonant inelastic x-ray scattering (RIXS) \cite{RIXS} and x-ray linear dichroism (XLD).

We also outline possible extensions of our minimal model. Beyond dipole injection, the framework can be directly applied to investigate quadrupole injection, for instance via the orbital-torsion Hall effect \cite{Han}. Moreover, the model can be generalized to magnetic materials with spin–orbit coupling by incorporating a term of the form $\boldsymbol{\sigma}\otimes\boldsymbol{L}$, where $\boldsymbol{\sigma}$ denotes the vector of Pauli matrices. This extension enables the exploration of interconversion phenomena among orbital (dipole and quadrupole), spin, and charge degrees of freedom. Within this generalized framework, coupled continuity equations for spin and orbital moments can be constructed to describe the dynamics of spin and orbital torques.


\textit{Acknowledgment---} C.S., D.G., Y.M., and A.M. were supported
by EIC Pathfinder OPEN grant 101129641 “OBELIX” and the France 2030 government investment plan managed by the French National Research Agency under grant reference PEPR SPIN – [SPINTHEORY] ANR-22-EXSP-0009 and [OXIMOR] ANR-24-EXSP-0011. J.L. was supported by the Research Council of Norway through Grant No. 353894 and its Centres of Excellence funding scheme Grant No. 262633 “QuSpin.” Y.M. gratefully acknowledges financial support from the Deutsche Forschungsgemeinschaft 
(DFG, German Research Foundation) - TRR 288/2 - 422213477 (project B06).


\bibliography{bib}

\end{document}